\documentclass[twocolumn,superscriptaddress,showpacs,amsfonts]{revtex4-1}
\pdfoutput=1

\usepackage{epsfig}
\usepackage{graphicx}
\usepackage{amsmath}
\usepackage{amssymb}
\usepackage{color}
\usepackage{bm}

\begin{document}

\title{Zigzag and Checkerboard Magnetic Patterns in Orbitally \\ Directional Double-Exchange Systems}

\author{W. Brzezicki}

\affiliation{CNR-SPIN and Dipartimento di Fisica
``E. R. Caianiello\textquotedblright{}, \\
 Universit\'a di Salerno, IT-84084 Fisciano (SA), Italy}
 
 \affiliation{Marian Smoluchowski Institute of Physics, Jagellonian University,
Reymonta 4, 30-059 Krak\'ow, Poland }

\author{C. Noce}
\affiliation{CNR-SPIN and Dipartimento di Fisica
``E. R. Caianiello\textquotedblright{}, \\
 Universit\'a di Salerno, IT-84084 Fisciano (SA), Italy}
 
\author{A. Romano}
\affiliation{CNR-SPIN and Dipartimento di Fisica
``E. R. Caianiello\textquotedblright{}, \\
 Universit\'a di Salerno, IT-84084 Fisciano (SA), Italy}
 
\author{M. Cuoco}
\affiliation{CNR-SPIN and Dipartimento di Fisica
``E. R. Caianiello\textquotedblright{}, \\
 Universit\'a di Salerno, IT-84084 Fisciano (SA), Italy}
 
\date{\today}

\begin{abstract}
We analyze a $t_{2g}$ double-exchange system where the orbital directionality of the 
itinerant degrees of freedom is a key dynamical feature that self-adjusts 
in response to doping and leads to a
phase diagram dominated by two classes of ground-states 
with zigzag and checkerboard patterns. 
The prevalence of distinct orderings is tied to the formation of orbital molecules that 
in one-dimensional paths make insulating zigzag states 
kinetically more favorable than metallic stripes, thus allowing for a novel
doping-induced metal-to-insulator transition. 
We find that the basic mechanism that controls the magnetic competition is the
breaking of orbital directionality through structural distortions and 
highlight the consequences of the interorbital Coulomb interaction. 
\end{abstract}

\maketitle

Transition metal (TM) oxides are fascinating materials characterized by a subtle interplay 
between charge, spin and orbital degrees of freedom, which in many cases gives rise to complex types 
of collective behavior.
Though firstly thought as a prerogative of $3d$ systems \cite{Ima98,Dagotto2005}, this class of phenomena seems 
now to be ubiquitous in $4d$ and $5d$ ones \cite{Kim2009,Pesin2010}. A key role in their occurrence is played on one side 
by the frustrated localized-itinerant nature of the magnetic correlations, and on the other side by the peculiar 
orbital dependent electron dynamics in partially filled $e_g$ and $t_{2g}$ sectors of $d$-shells.
Prototype examples of electronic self-organization 
are provided by the magnetic and charge orders  
detected in layered manganites \cite{Moritomo,Sternlieb,Tokura} and nickelates \cite{Tranquada}.

The formation of spin-charge density modulations is strongly related to the
orbital character of the electronic system as demonstrated by the dominant role
of lattice distortions in itinerant $e_g$ systems \cite{Hotta2001,Hotta2004,Dong2009} 
compared with the spin-orbital exchanges in models of
insulating $t_{2g}$ electrons \cite{Kruger2009,Wrobel2010}.
More unexplored is the case of partially localized 
$t_{2g}$ electrons in sistems with low dimensionality and competing magnetic correlations.
In this context, new phenomena have recently been observed and investigated in hybrid oxides 
with partial substitution of inequivalent TM ions \cite{Qi2012,Cao2014,Lei2014,Dhital2014,Brzezicki2015}. 
Particularly fascinating are the Mn-doped layered Sr-ruthenates that represent a paradigmatic example 
of non-trivial coupling between itinerant ferromagnetic (FM) and 
localized antiferromagnetic (AF) degrees of freedom \cite{Mat2005,Ortmann2013} in the 
doped $t_{2g}$ sector with a resulting 
metal-insulator transition (MIT) \cite{Mat2005,Hu2011,Hos12,Pan2011} 
and magnetic order \cite{Mat2005,Mes12,Hos12,Hos2013,Ortmann2013,Li2013} that are decoupled and robust 
over a large range of doping \cite{Hu2011,Hos12,Hos2013}. 

\begin{figure}[ht]
\includegraphics[width=0.6\columnwidth]{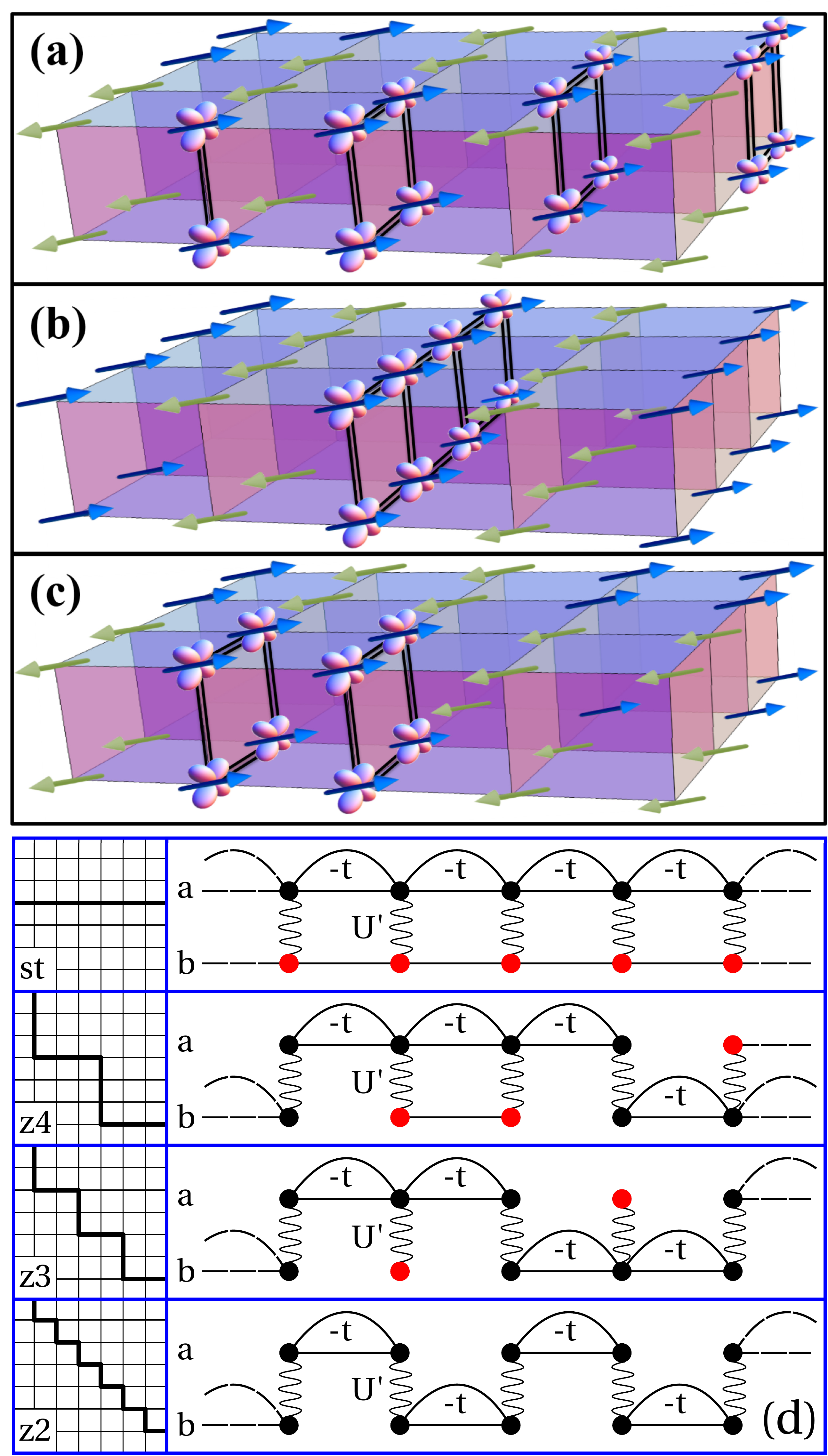}
\caption{(color online). Schematic view of layered $t_{2g}$ systems with $d_{xz/yz}$ itinerant degrees of freedom and 
different antiferromagnetic and orbital patterns: 
(a) two-site segment zigzag $z2$ (E-phase), (b) straight stripes $st$, and (c) $2\times 2$ - cell checkerboard $c2$. 
Doubled lines denote the constrained hopping for $d_{xz/yz}$ orbitals in undistorted system.
(d) sketch of the hoppings for embedded 1D paths (thick lines);
stripe $st$ and zigzag $z4$, $z3$ and $z2$ phases.
For each site there are two orbital flavors (dots) $a$
and $b$, black (red) color marks active (inactive) orbitals. 
The arcs mark bonds where hopping is allowed.The structure is repeated in the second layer. 
The wiggly line denotes the onsite interorbital coupling $U^{\prime}$.
\label{fig:fig1}}
\end{figure}

In this Letter, we show general features of orbitally directional double-exchange (DE) layered systems 
as a novel metal-to-insulator transition and two predominant types of orderings within the phase diagram.
DE mechanism is known to be at the origin of itinerant ferromagnetism in $e_g$ systems and, 
when the 
superexchange between localized spins is considered, to yield exotic 
magnetic structures~\cite{DagottoReview,Yunoki2000,Hotta2001,Hotta2003,Kumar2010,Kumar2010,Garcia2000,Garcia2002,GarciaPRL2004,Aliaga2001} 
and other states based on electronic self-organization~\cite{Dong2009,Ye2009},
whose stability often relies on additional microscopic couplings, and is confined to
specific electron densities.  
In the orbitally directional DE system, 
we show that the formation of orbital molecules, within one dimensional (1D) FM configurations, 
is crucial to have insulating zigzag patterns that are
energetically more favorable than metallic straight stripes, thus allowing for a novel kind of MIT.
We find that, due to the orbital directionality, the competition between AF and FM correlations in layered systems
makes antiferromagnetically coupled FM zigzag stripes and checkerboard clusters (Fig. \ref{fig:fig1}) the dominant 
patterns in the phase diagram over a large range of doping. 
We demonstrate how the breaking of the orbital directionality as
well as the inclusion of the Coulomb interaction can significantly affect
the zigzag-checkerboard competition and lead to orbital/charge ordering in the ground-state.

The model Hamiltonian is
\maketitle
\begin{eqnarray}
{\cal H} & = & \sum_{i,\sigma}\sum_{{\alpha,\beta=a,b\atop {\hat{\gamma}}={\hat{a} ,\hat{b} ,\hat{c} }}}t_{\hat{\gamma},\alpha \beta} \left(d_{i,\alpha\sigma}^{\dagger}d_{i+\hat{\gamma},\beta\sigma}+h.c.\right)\nonumber \\
 & - & J_{H}\!\!\!\sum_{i,\alpha=a,b}\!\!{\bf s}_{i\alpha}\cdot{\bf S}_{i}+J_{AF}\!\sum_{i,{\hat{\gamma}=\hat{a},\hat{b}}}{\bf S}_{i}\!\cdot\!{\bf S}_{i+{\hat{\gamma}}} \label{eq:Ham} +
U^{\prime}\!\sum_{i} n_{i,a} n_{i,b} \nonumber \,,
\end{eqnarray}
where $d^{\dagger}_{i,\alpha\sigma}$ is the electron creation operator at the site $i$ with spin $\sigma$ for the orbital
$\alpha$. For convenience $(a,b,c)$ indicate the $(yz,xz,xy)$ orbitals which
are perpendicular to the corresponding
bond direction, with $\hat{a}$, $\hat{b}$, and $\hat{c}$ being the unit vectors along the lattice symmetry directions.
$n_{i,\alpha}$ is the local electron density for the orbital
$\alpha$, ${\bf s}_{i\alpha}\!=\!\frac{1}{2} d_{i,\alpha,m}^{\dagger}\vec{\sigma}_{\!m,m}d_{i,\mu,n}$
and ${\bf S}_{i}$ denote the spins for the $d_{xz/yz}$ and
$d_{xy}$ orbitals, respectively.  
$t_{\hat{\gamma},\alpha \beta}$ is the nearest-neighbor hopping amplitude between the orbitals $\alpha$ and $\beta$
for the bond along the $\hat{\gamma}$. 
We take the tetragonal amplitudes $t_{\hat{a},bb}=t_{\hat{b},aa}=-t$ with $t$ as energy scale unit. 
$J_{H}$ stands for the Hund coupling between localized and itinerant electrons while $J_{AF}$ is the
AF superexchange of $c$ orbitals.
The hole doping $x$ leads to $d_{xz/yz}^3$-$d_{xz/yz}^2$ partial substitution.
We consider that $J_{AF}$ follows from virtual charge excitations in the presence of strong on-site Coulomb interaction and we focus 
on the Hund coupling and the interorbital Coulomb interaction $U^{\prime}$ for the xz/yz orbitals. This assumption is also 
motivated by the connection between DE and orbital-selective-Mott physics~\cite{Medici2011,Biermann2005,Rincon2014}.
To determine the ground-state (GS) the local spins
are considered as Ising variables \cite{supmat1}.

\begin{figure}
\includegraphics[width=1\columnwidth]{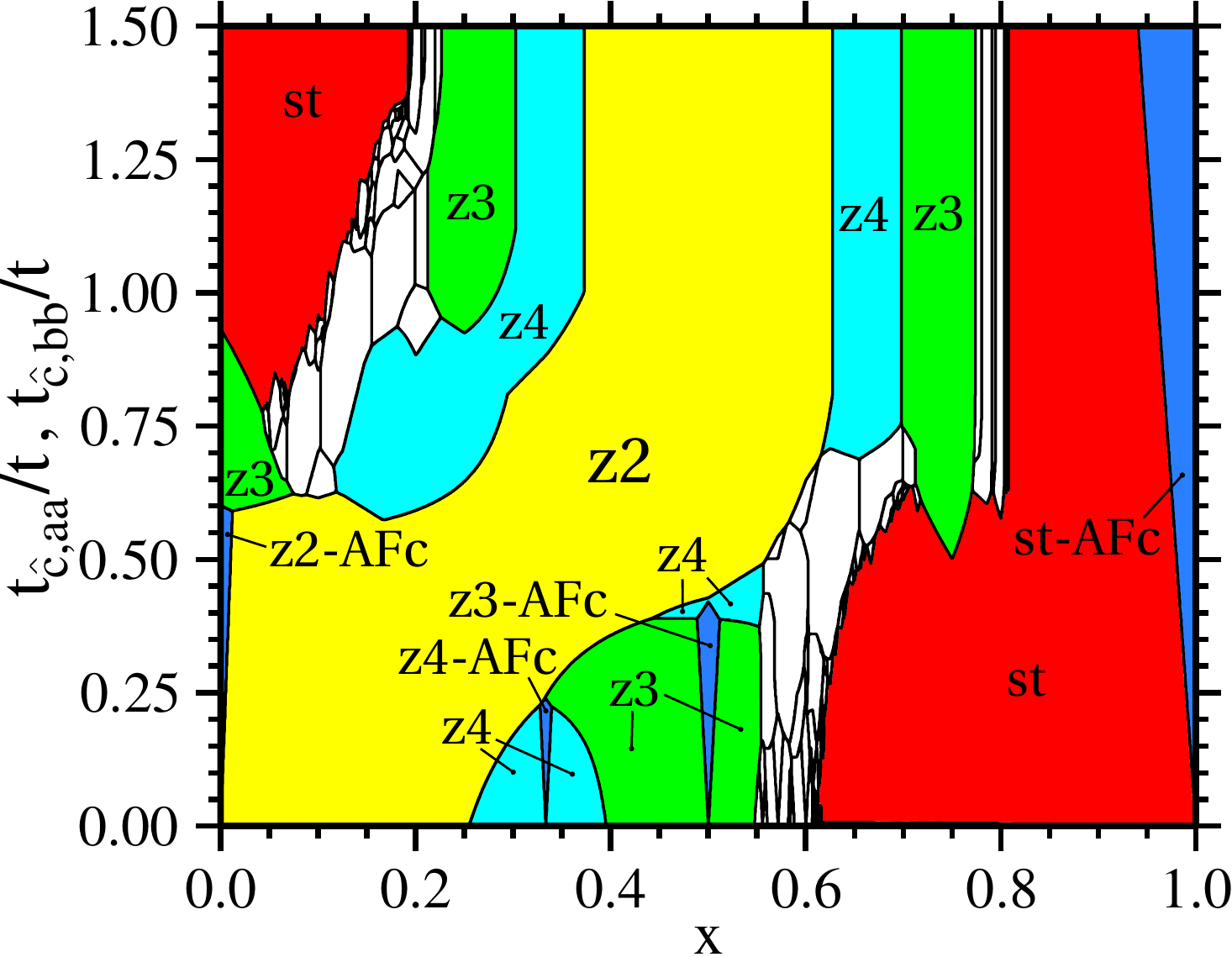}\protect\caption{(color online).
Ground state diagram for embedded 1D FM patterns as a function of the hole doping $x$ and the $c$-axis hoppings $t_{\hat{c},aa}$ and $t_{\hat{c},bb}$ at $U^{\prime}=0$.
AF$c$ denotes $c$-axis AF alignment. Unlabeled regions denote $zn$ zig-zag phases with $n>4$.}
\label{fig:fig3}
\end{figure}

We start by dealing with isolated 1D-FM paths in a two-dimensional (2D) layered structure for 
the undistorted noninteracting case ($U^{\prime}=0)$.
The model (\ref{eq:Ham}) is solved for all the allowed 1D configurations made of straight stripe ($st$) and zigzag patterns (z$n$) 
with $n$-atom 
segments (Fig.~\ref{fig:fig1}(d)).
In the $st$ case, one orbital is blocked and the other one is itinerant, while 
the zigzag has both orbitals active along the corresponding segments.
The $d_{xz/yz}$ connectivity for the zigzag profiles can be mapped on a 1D view as reported in 
Fig.~\ref{fig:fig1}(d).
For any zigzag, the GS
is insulating and it factorizes in product of orbital active
electronic configurations within each segment.
Due to the freedom of one or two itinerant channels, the issue is to determine which
path lowers the kinetic energy when doping the system.
In Fig.~\ref{fig:fig3} we report the GS diagram as a function of doping and interlayer hopping $t_{\hat{c},aa/bb}$. 
We note that below half-filling ($x<0.5$) the $z2$ zigzag pattern is the dominant state, 
reflecting the general tendency, induced by doping, to avoid electron propagation along straight stripes.

To get more insight, it is instructive to compare the GS energy of $z2$ and $st$ at $x=0$ and for the monolayer case.
The $st$-path has a total energy per site equal to $-4 t/(2 \pi)$.
On the other hand, the $z2$ ground state is made of disconnected two-atom clusters with filled bonding and empty antibonding
configurations. Hence, the GS energy per site is equal to $-t$ and it is lower than that of the
$st$ configuration. The robustness of $z2$ relies on the
possibility to fill only bonding states and to avoid configurations with nodes in the
confined segments. 
We argue that the $st$ state is generally unstable towards the formation of a molecular configuration where electrons {\it condense} 
in the lowest energy state and tend to minimize the nodes in the quantum wave-function.

A striking feature of the 1D diagram is that the doping-induced transition from metallic $st$ to
insulating $z2$ state goes through many intermediate zigzag configurations having long unit segments.
This represents an unusual type of MIT with a breakdown of metallic paths into zigzag insulating ones.
When considering the bilayer system, the flat orbitals acquire itinerancy along the interlayer direction (Fig.~\ref{fig:fig1}(a)) 
and compete with the in-plane $z2$ bonding.
The result is the stabilization, at low and high doping, of larger zigzag configurations ($z3$ and $z4$) as well as straight stripes.
We find that a change of the hopping amplitude around the doped site
does not affect much the phase boundaries mostly due to the robust insulating character of the z$n$ states.

\begin{figure}[t]
\includegraphics[clip,width=0.85\columnwidth]{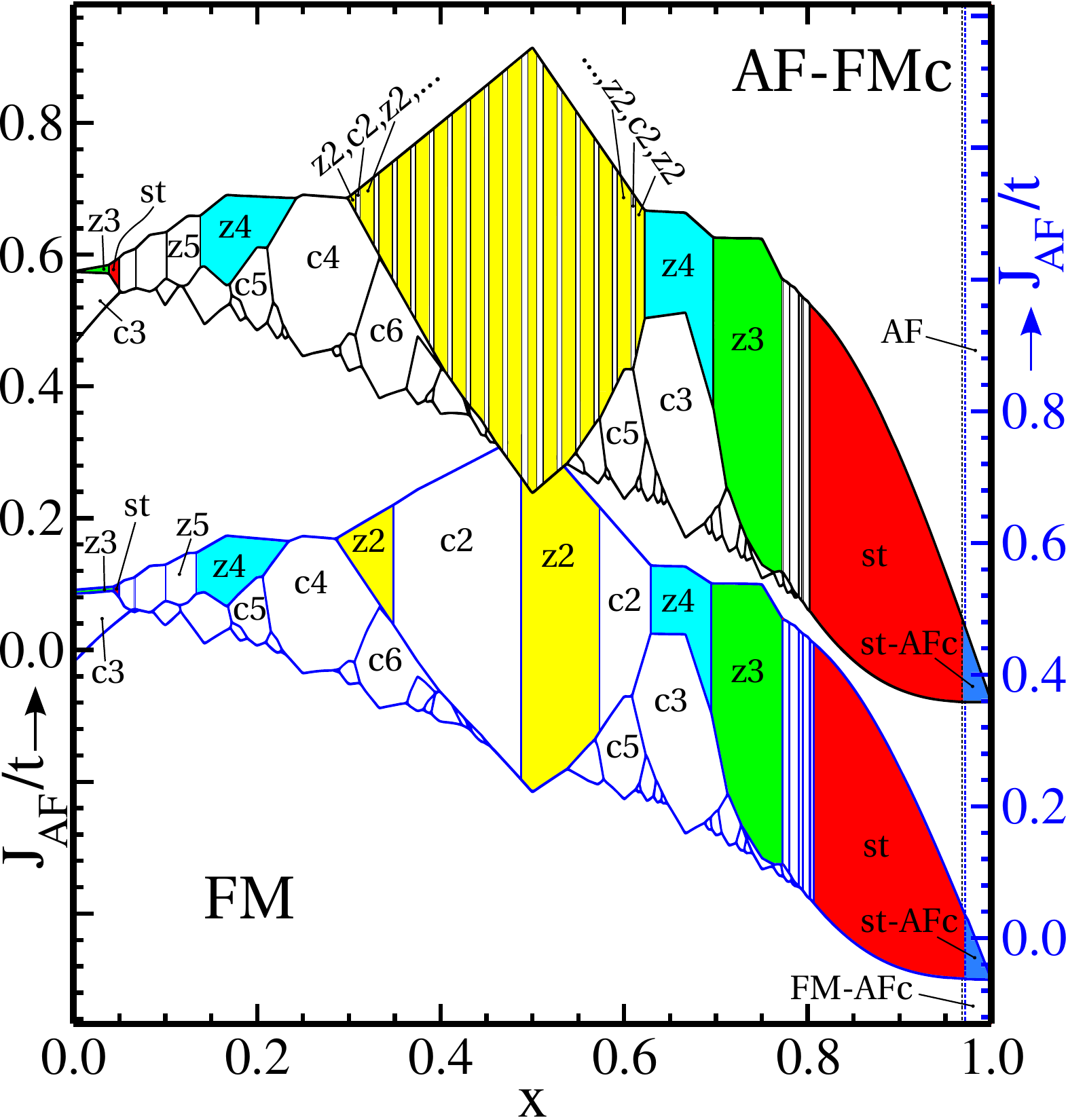}
\protect
\caption{(color online). Phase diagram of the 2D $t_{2g}$-DE model for the bilayered system vs doping $x$ and $J_{AF}$,
with no distortions, $U^{\prime}=0$ and in the presence of octahedral tilting angle $\theta$: $\theta=0^{\circ}$ 
(top side plot, left vertical scale) and $\theta=10^{\circ}$ (bottom side plot,
right vertical scale). 
FM and AF-FMc denote FM/AF layers with FM $c$-axis alignment. 
Unlabeled regions are z$n$ and c$n$ phases with $n>6$.
The parameters used are at $\theta=0$: $J_H=100 t$ and $t_{\hat{a},bb}/t_{\hat{c},aa(bb)}$=0.8 $t$ (for $\theta=10^{\circ}$ see Ref. \onlinecite{supmat1}).
\label{fig:fig4}}
\end{figure}

In order to address the role of dimensionality 
we consider a double-layered system 
for a representative value of tetragonal anisotropy, i.e. $t_{\hat{a},bb}/t_{\hat{c},aa(bb)}$=0.8 $t$. 
The results are obtained assuming the Hund coupling as the dominant energy scale, i.e. $J_{H}=100 t$. 
Smaller values of $J_{H}$ (i.e. $J_{H}=10\,t$ and $5\,t$) have also been considered and modify the diagram by reducing the area of stability of large
size z$n$ and c$n$ patterns, slightly shifting the boundaries of the $z2$ and $c2$ phases.
Search for spin patterns has been also performed in representative points of the diagram by means of Montecarlo simulations 
employing the Metropolis algorithm.
This analysis confirms the zigzag and checkerboard phases as the dominant ones.
As shown in Fig.~\ref{fig:fig4}, FM and AF-FMc states occur at small and large $J_{AF}$ amplitudes. The AF-FMc is made of
AF layers coupled ferromagnetically. The high doping 
regime, i.e. $x>0.8$, exhibits a major tendency towards $st$ phase (Fig. \ref{fig:fig1}(b)), 
which may be relevant for layered 
systems with
hole doped Mn$^{4+}$ manganites or vanadates~\cite{Miyasaka2000,Ishihara2005,Wohlfeld2006}.
Since the $d_{xy}$ band does not hybridize along the $c$-direction, 
the effective AF coupling is vanishing and the interlayer ferromagnetism is always favored, except close to $x=1$ due to the Pauli principle.
As expected from the 1D study, the $z2$ zigzag states (Fig.~\ref{fig:fig1}(a)), are energetically favorable. 
However, they are quasi degenerate with checkerboard states (c$n$), made of
$n\times n$ FM clusters that are coupled antiferromagnetically (Fig.~\ref{fig:fig1}(c) for the $c2$ configuration).
This result primarily arises from the fact the $z2$ and $c2$ electronic spectra are identical in the unit block due to the $t_{2g}$ orbital directionality.
The coupling between AF spin domains brings corrections of the order of $1/J_{H}$ in the dispersion and 
is responsible for the degeneracy removal and the cascade of doping induced transitions between $z2$ and $c2$.
%
\begin{figure}[t]
\includegraphics[clip,width=1.0\columnwidth]{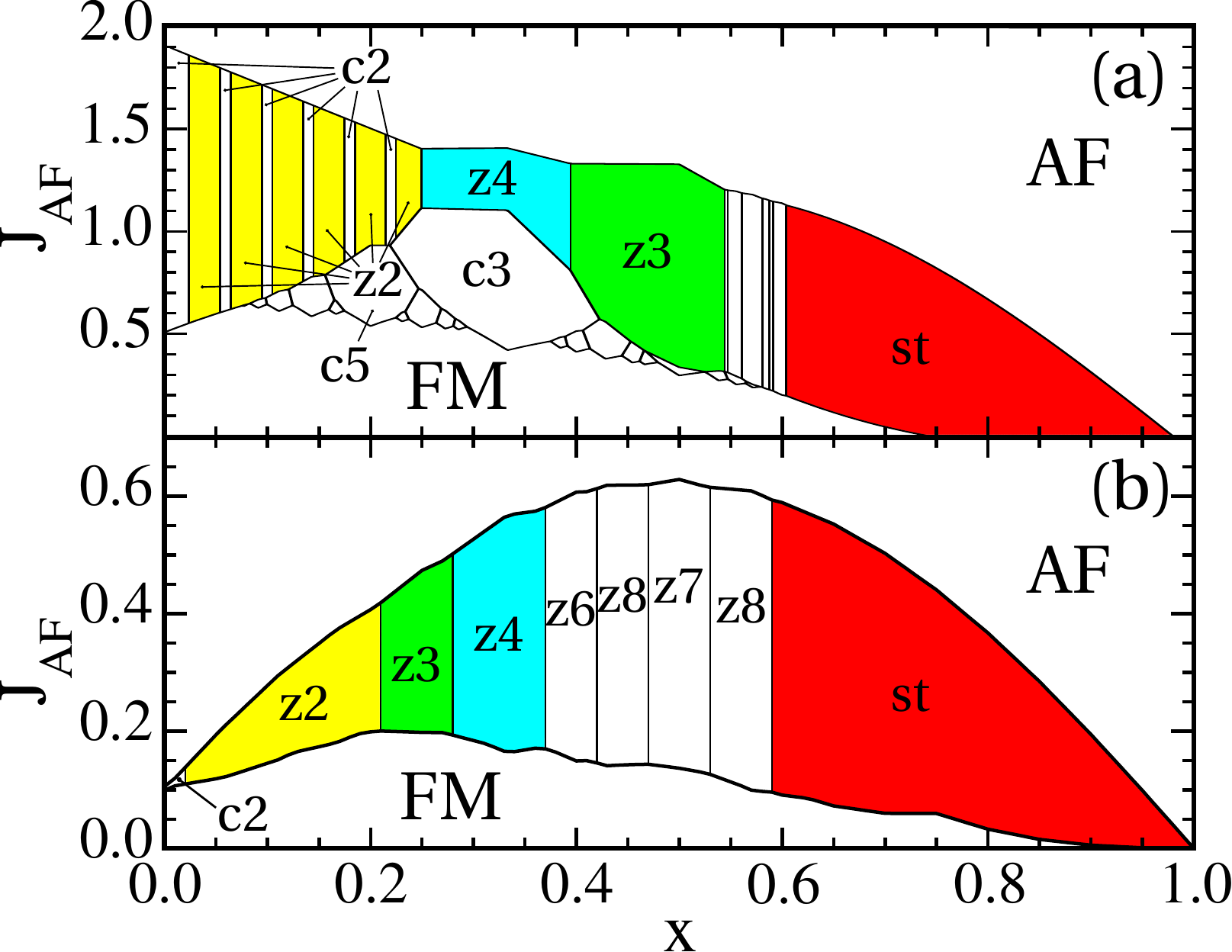}
\protect
\caption{(color online). Phase diagram of the monolayer tetragonal system with $U^{\prime}=0$ (a) and $U^{\prime}=10\,t$ (b).}
\label{fig:fig5}
\end{figure}
%
Rotation and tilting of the octahedra with respect to the $c$-axis are the main processes that 
break the orbital directionality of the $d_{xz/yz}$ orbitals.  
To study their consequences we have employed a general approach based on the Slater-Koster rules~\cite{SlaterKoster,supmat1} 
The rotation does not modify the phase diagram boundaries because the hopping matrix transformation 
for the $d_{xz/yz}$ can be gauged away in the kinetic term.
Then, it can only lead to transitions via a renormalization of $J_{AF}$.
The tilting mixes $d_{xz/yz}$ orbitals \cite{supmat1} and 
thus breaks the directional disconnection resulting in the removal of the quasi-degeneracy between the zigzag and checkerboard states (Fig.~\ref{fig:fig4}). 
The effects are more evident close to $x=0.5$ where tilting favors the $c2$ state. Moreover, by reducing the interlayer diagonal hopping it
makes $z2$ appearing at a lower doping down to $x\sim 0.3$ as in the 1D case. 
 
We now discuss the role played by the $U^{\prime}$ interorbital Coulomb interaction~\cite{Held2000, Rozenberg1998, Golosov2010}. 
We notice that $U^{\prime}$ is manifestly distinct for 
the zigzag and 
checkerboard patterns as due to the interplay of the pattern configuration
and $t_{2g}$ directionality. In the checkerboards $U^{\prime}$  
frustrates the confined $d_{xz/yz}$ charge motion in 
the spin-polarized blocks.
On the contrary, in the zigzag there is an explicit difference for $U^{\prime}$ 
at the inner and the corner sites (Fig. \ref{fig:fig1}). 
To deepen such comparison we
focus on the tetragonal monolayer system. Inside the
zigzag units one of
the two orbitals is localized and the charge degree of freedom behaves 
like a classical variable. Then,
the intra-segment interacting problem maps into the 1D Falicov-Kimball model \cite{Falicov1969} 
whose itinerant spinless electrons in the active $d_{xz/yz}$ orbitals locally couple to a classical variable
describing the density of the inactive ones. 
At the corner sites, it is not possible to have a classical behavior and $U^{\prime}$ is
fully quantum
as the orbital electron density varies between zero and one 
depending on the intra-segment electronic configuration. 
The emerging low energy scenario is particularly clear for the z2 state.
One can show that the interacting problem 
can be exactly mapped into an Ising model in a transverse field.
Then, the GS is made of 
domain walls that propagate along the
zigzag path controlled by the ratio $U^{\prime}/t$. 
Such modes generally allow for a kinetic energy gain. 
For larger zigzag patterns the inner-corner electronic separation leads to
enhanced electron-hole correlations at the corner to avoid $U^{\prime}$
and the GS exhibits a tendency to an asymmetric charge distribution 
inside the zigzag block. 

In order to quantitatively account for the role of $U^{\prime}$ we employ an exact diagonalization 
study based on the Lanczos algorithm, simulating both zigzag and checkerboard patterns with 
different cluster size \cite{supmat1}. 
In Fig. \ref{fig:fig5} we report the phase diagram for the 
2D monolayer tetragonal system at $U^{\prime}$=0 (a) and $U^{\prime}=10\,t$ (b).
A general outcome is that the intrablock kinetic energy is reduced by 
$U^{\prime}$ and increases its competition with the AF exchange. 
Such aspect is particularly relevant for the stability of the checkerboard states as 
the AF energy contribution cancels out for all the 
zigzag states while, except for c2, 
it is detrimental for the checkerboard configurations.
Then, the 
window of stability of the zigzag and checkerboard states shrinks in terms of 
the $J_{AF}/t$ ratio. 
At low doping the competition is purely of electronic origin as the 
AF exchange is equivalent for the z2 and c2 states. 
It is worth pointing out how $U^{\prime}$ drives the stability of the z2 state.
$U^{\prime}$ tends to kinetically frustrate the electron propagation for both z2 and c2 
and this effect is quite strong at low doping. However, for z2 such constraint 
can be released by the propagation of the interorbital defects along the
zigzag path. Such collective behavior
is absent in the checkerboard c2 configuration. 
Approaching the range of doping where in the noninteracting limit longer zigzag and checkerboard 
states compete (e.g. z3, z4 and c3), we observe that $U^{\prime}$ favors the zigzag phases. 
This result is a cooperative effect between $U^{\prime}$ and $J_{AF}$ because the
interaction renormalizes the kinetic energy and then the AF
exchange can easily overcome the difference in the electronic contribution that makes the checkerboard patterns more 
favorable in the range of doping 
close x=0.5. 
Above half-filling, the density of minority spins is quite dilute and the Coulomb interaction is not much influent.
Hence, the Coulomb interaction confirms the occurrence of a doping induced MIT moving from a
very dilute metallic stripe ($x\sim 1$) to intermediate long zigzag ($x\sim 0.5$) and dense short zigzag patterns ($x\sim 0$) 
that is akin to an electronic gas-to-liquid-to-crystal changeover. 
A distinct feature occurs when considering the charge profile of the
zigzag GS. 
We find that, among all the zigzags, z3 state exhibits a GS with a charge density wave 
with nonuniform electron density. 
The correlated z3 phase has a charge and orbital ordering that
remarkably can yield a nonvanishing electric dipole in each zizag unit making the 
single zigzag chain prone to a ferroelectric instability. 

In summary, we have determined the spin-charge-orbital modulated patterns 
that naturally emerge in orbitally directional $t_{2g}$ DE systems 
for an extended range of doping and couplings. 
We argue that these results can be of high relevance for understanding the phase diagram of
Mn-doped layered Sr-ruthenates or other oxides where TM substitutions can lead to $d^3$-$d^4$ or $d^2$-$d^3$ charge doping 
in the $t_{2g}$ sector with a partial localization of one orbital degree of freedom.

We thank A.M. Ole\'s and R. Jin for valuable discussions.
W.B. acknowledges support by the Polish National
Science Center (NCN), Project No. 2012/04/A/ST3/00331,
and the Foundation for Polish Science (FNP) within the START program.

\section{Supplemental Material}

In this section we present the parameterization used for the hopping amplitudes between the $d_{xz}$ and $d_{yz}$orbitals   
in the presence of octahedral rotation and tilting with respect to the $c$-axis.
Furthermore, we provide extra details about the methodology
adopted for determining the phase diagram of the $t_{2g}$ orbitally degenerate double-exchange model including the case
with interorbital Coulomb interaction. 

In order to analyze the role played by octahedral distortions, we have determined the modification of the hopping amplitudes between the orbitals involved on a distorted bond 
by means of the Slater-Koster rules \cite{SlaterKoster}.
Since the integrals that appear in the hopping term rapidly decrease as a function of the distance between the two selected atoms, 
one can restrict the analysis to nearest neighbor sites.
Let us then consider a single two-center TM$_1$-O (transition metal-oxygen) bond with TM$_1$ located at $R(\rm{TM_1})$ and O at $R(\rm{O})$, respectively. 
Following Ref. \cite{SlaterKoster}, we take Wannier orbitals with respect to a set of
rectangular axes along the undistorted TM$_1$-O bonds. Then, the integrals between $p_i$ $(i=x,y,z)$ and d$_{xy/xz/yz}$ orbitals are parameterized in terms of the 
overlap amplitudes between the $pd\pi$ and $pp\sigma$ ones and the distance between the 
pairs of sites is taken into account with respect to the TM1-O angle. 
For the latter, the direction
cosines of the vector $R(\rm{O})-R(\rm{TM_1})$, pointing
from one atom to the other, are given by $(l, m, n)$ with respect to the symmetry axes.
The values of the needed matrix elements can be taken directly from the table in Ref. \cite{SlaterKoster}.
For instance, the integrals between two representatives cases of $p$- and $d_{xy/xz}$-orbitals which are of interest for the
model considered here can be expressed as 
\begin{eqnarray*}
E_{p_x,d_{xy}}(l,m,n)&&=\sqrt{3} l^2 m (pd\sigma)+ m (1-2 l^2) (pd\pi) \nonumber \\
E_{p_z,d_{xz}}(l,m,n)&&=\sqrt{3} l^2 m (pd\sigma)+ l (1-2 n^2) (pd\pi) \,
\end{eqnarray*}
with $(pd\sigma)$ [$(pd\pi)$] being the overlap integral between the $d\sigma$ and $p\sigma$ [$d\pi$ and $p\pi$] orbitals.
These integrals generally depend on the distance between the two considered atoms.
Similar expressions can be obtained for all the $p_i$ $(i=x,y,z)$ and $d_{xy/xz/yz}$ orbitals \cite{SlaterKoster}.
\begin{figure}[t]
\includegraphics[clip,width=0.8\columnwidth]{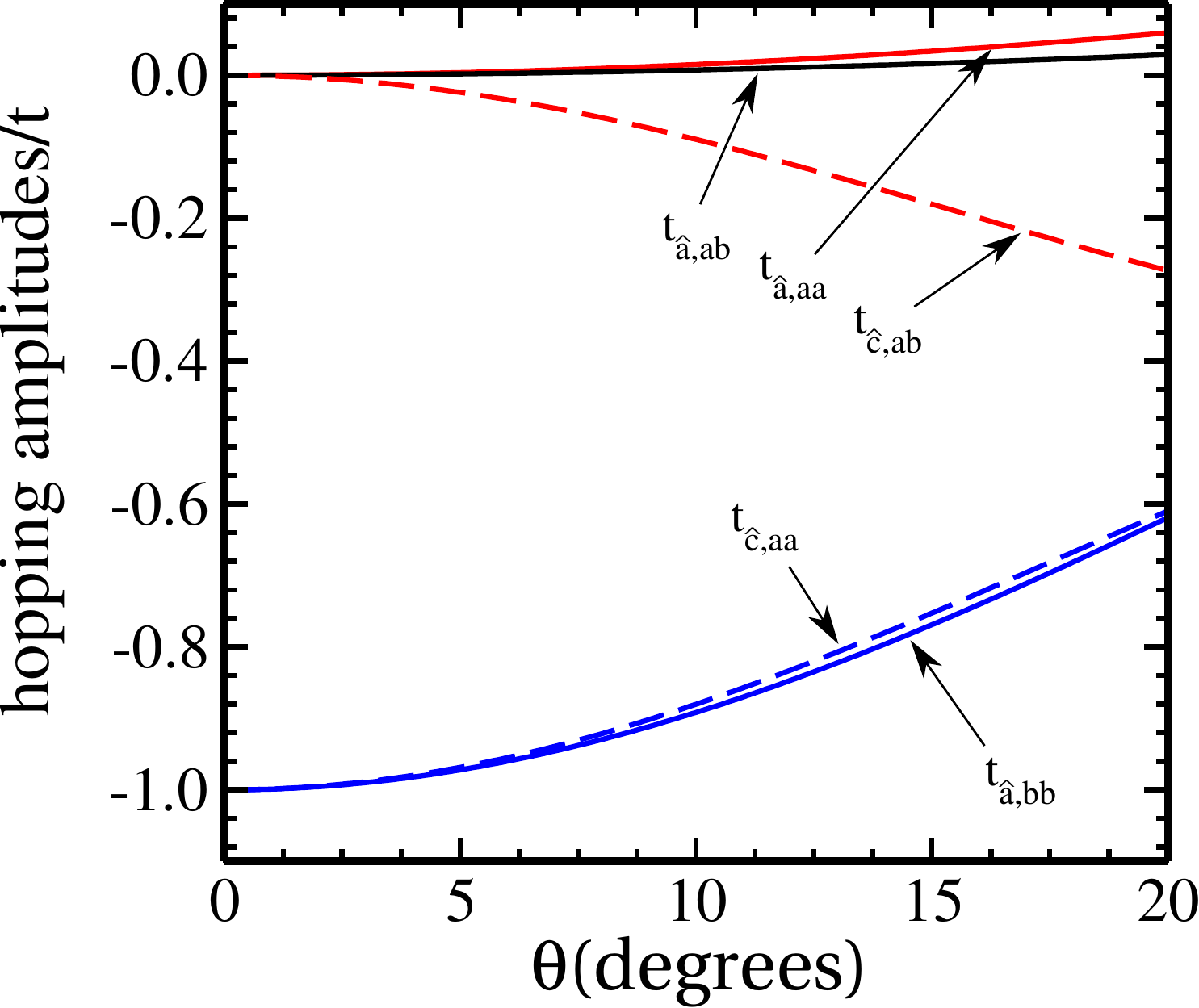}\protect\caption{
Evolution of the effective hopping amplitudes between the $a$ and $b$ orbitals as a function of the tilting angle $\theta$ along the $\hat{a}$,$\hat{b}$, and $\hat{c}$
symmetry directions.
\label{fig:hop}}
\end{figure}
Using these amplitudes, we can determine for the undistorted bond the effective hopping between neighboring transition metal ions. 
As in the manuscript, we employ $(a,b,c)$ to indicate the $(yz,xz,xy)$ orbitals which
are perpendicular to the corresponding
bond direction with $\hat{a}$, $\hat{b}$, and $\hat{c}$ being the unit vectors along the lattice symmetry directions.
Then, the effective hopping between the $b$ orbitals along the $\hat{a}$ axis is given by
\begin{eqnarray}
t_{\hat{a},bb}=E_{p_z,d_{xz}}(1,0,0) \times E_{p_z,d_{xz}}(-1,0,0)=-(pd\pi)^2 \,
\end{eqnarray}
\noindent and the same is for $t_{\hat{b},aa}$. In the absence of distortions it is immediate to 
verify that $t_{\hat{a},ab}=t_{\hat{b},ab}=0$. For the computational analysis, we assume the ratio of the overlap 
integrals to be $(pd\sigma)/(pd\pi)=2.5$ that is of an order of magnitude that is common to layered oxides. 
Then, for the undistorted case we set $t_{\hat{a},bb}=t_{\hat{b},aa}=-t$ for the in-plane hopping 
with $t$ being our unit of energy. For the $e_g$-$d_{x^2-y^2}$ orbitals a similar calculation gives a tetragonal hopping equal to $-3/4(pd\sigma)^2$.  
Let us consider the expression of the effective hopping when the octahedra has a tilting $\theta$ with respect to the $c$-axis. 
In this configuration, the M1-O-M2 bond angle is different from $180^\circ$ and thus more than one cosine direction enters
in the effective hopping calculation.
The result for the in-plane hopping can be expressed as 
\begin{eqnarray}
t_{\hat{a},bb}=\sum_{i} E_{p_i,xz}(l,m,0) \times E_{p_i,xz}(-l,m,0) 
\end{eqnarray}
where $l$ and $m$ are proportional to $\cos(\theta)$ and $\sin(\theta)$, respectively.
Similar expressions can be obtained for the rotation of the octahedra and for the effective hoppings along the other M-O bond directions.
In Fig. \ref{fig:hop} we have reported the evolution of the hopping amplitudes along the various symmetry directions involving the $a$ and $b$ orbitals 
scaled to the undistorted unit $t$. As one can note, 
the tilting of the octahedra mainly affects the off-diagonal hopping along the $c$-axis, 
i.e. t$_{{\hat{c}},ab}$, which grows from zero up to about $t_{\hat{a},bb}/3$ at $\theta\sim 20^{\circ}$. 
With respect to the undistorted configuration the diagonal hopping matrix elements exhibit a reduction of 40$\%$ when, 
for instance, $\theta$ varies between 0 and 20$^{\circ}$. The tilt considered in this paper is staggered 
and has a four-sublattice structure differing by the sign of the tilt angle or the tilt axis being either $\hat{a}+\hat{b}$
or $\hat{a}-\hat{b}$. This does not affect the absulute values of the hopping amplitudes on different bonds but 
induces alternating signs. In this case however the alternation is such that it can be easily canceled by a proper gauge
transformation of the fermion operators on a chosen sublattice.

\begin{figure}[t]
\includegraphics[clip,width=0.8\columnwidth]{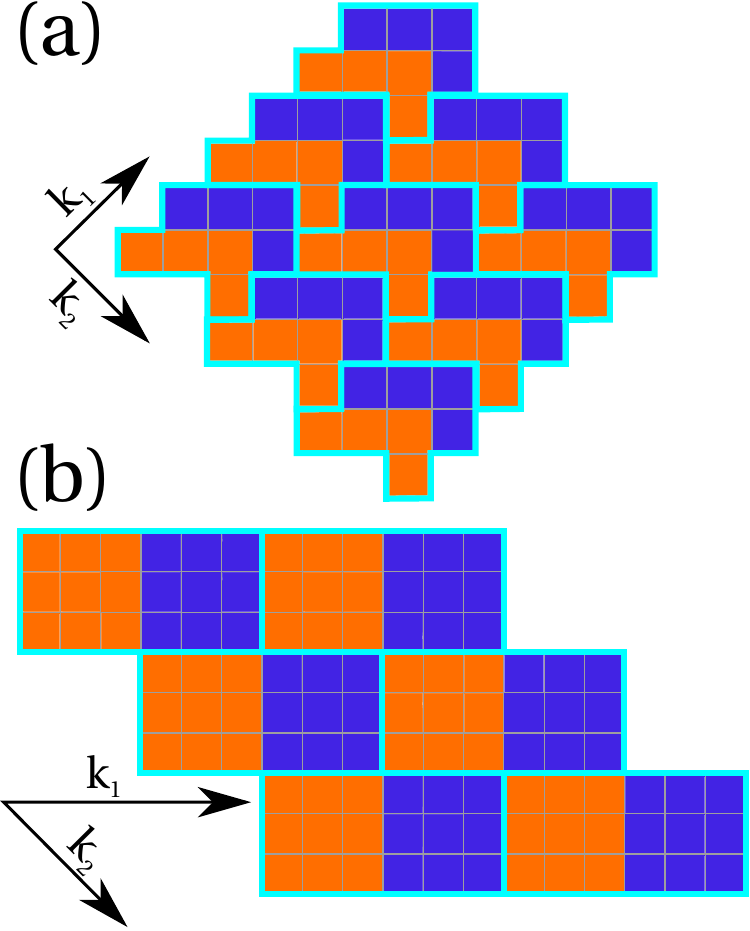}
\protect
\caption{
(color online). Schematic view of the lattice and the elementary cell used for the calculation in the reciprocal $k$-space of the 
(a) z3 phase and (b) c3 phase, respectively. Different colors of the small squares denote sites with
up and down spin polarization. The elementary cells are marked with thick lines with bright blue color.
All the unit cells contain a second layer for the bilayered structure having either the same or an opposite
spin polarization.
\label{zc_cells}}
\end{figure}

Since there are no qualitative differences with the results presented in the main text, 
we report here for completeness the modification of the phase diagram at $U^{\prime}=0$ 
in the presence of another channel of electronic itinerancy that is isotropic in the $ab$ plane. 
This type of process can mimic
the valence fluctuation at the site of the transition metal atom, for instance induced by 
substitutional doping, by adding an e$_{g}$-$d_{x^2-y^2}$ orbital to the DE model.
The selection of the e$_{g}$-$d_{x^2-y^2}$ orbital would be dictated by the octahedral 
distortions.
Due to the possibility of $d_{x^2-y^2}$ electrons to propagate along the whole zigzag path, 
there is a gain in kinetic energy that removes the degeneracy 
between the $z2$ and $c2$ configurations with a dominance of the zigzag state in the doping range of about $[0.25,0.5]$ (Fig.~\ref{fig:fig5}). 
The $st$ phase region almost disappears and checkerboard states with large domain sizes can get stabilized in the low and high doping region of the diagram. 
The AF energy scale is modified here as it reflects the difference in the amplitude of the $pd\sigma$ and $pd\pi$ hybridizations

\begin{figure}[t]
\includegraphics[clip,width=0.85\columnwidth]{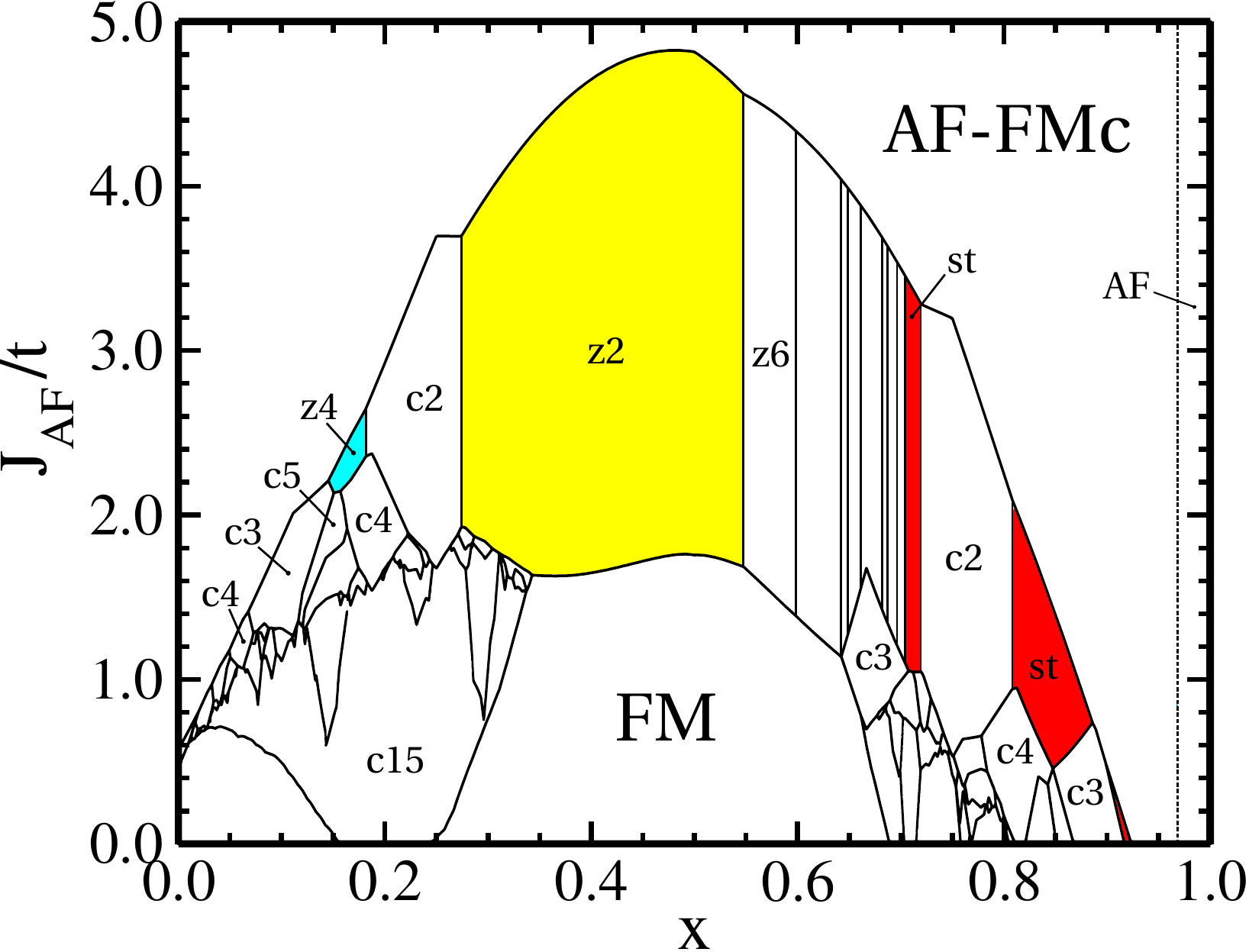}
\protect
\caption{(color online). Phase diagram at $U^{\prime}=0$ with no distortions and including the e$_{g}$-$d_{x^2-y^2}$ in the $t_{2g}$ DE-model. 
Unlabeled regions denote phases with $n>4$.
\label{fig:fig5}}
\end{figure}

Let us discuss the methodology applied for the determination of the one dimensional (1D) and two dimensional (2D) phase 
diagram of the $t_{2g}$ double-exchange model (Eq.1 in the main text). We start from the noninteracting case at $U^{\prime}=0$.
For the diagonalization of the matrix Hamiltonian in the reciprocal $k$-space we have used appropriate unit cells to
reproduce the zig-zag and checkerboard patterns (Fig. \ref{zc_cells}). The simulation has been performed for a finite size system by
having $N_1$ elements
in the direction of the basis vector $k_1$ and $N_2$ along $k_2$. 
These cells are always selected in such
a way that the whole lattice is approximately a square of the size $L\times L$ where
$L$, for the computation, is typically taken to be equal to $200$ sites for the case of periodic boundary conditions.
A special case is the one of the undistorted lattice where due to special symmetry of the matrix Hamiltonian a simulation with 
$L$ up to $1000$ sites can be easily achieved. 
In order to test the dependence of the results on the size of the lattice, clusters with up to $L=400$ sites have been
employed. 
The diagonalization of the Hamiltonian within a single unit cell at given $k_1$ and $k_2$ is
done by a canonical transformation and then the ground state energy is obtained by filling the energy 
levels up to the Fermi level according to the doping concentration $x$. 
For the undistorted case the canonical transformation involves only
sites lying along one line. Thus, it is possible to study checkerboard phases with very large unit cells 
since the
matrices that have to be diagonalized are of the order of $4n\times 4n$ where $n$ is the 
size of a single FM domain. Concerning the case in the presence of distortion, such transformation involves a matrix of size 
$8n^2\times 8n^2$ which significantly limits the maximal $n$ that can be computed.
In the case of the 1D patterns when only the $zn$ and the $st$ phases are considered, we use a very similar 
unit cell as the one shown in Fig. \ref{zc_cells} (a), but with only one zig-zag segment inside the cell corresponding to
a specific spin polarization. For such circumstance, the simulation is performed with number of cells $N_2$ along the $k_2$ direction 
in such a way that the total length of the chain is not smaller than $L=1000$.

Finally, we provide further technical details about the methodology and the approach employed to 
determine the ground state diagram when the interorbital Coulomb interaction is added. 
We use an exact diagonalization 
study based on the Lanczos algorithm and we simulate zigzag and checkerboard patterns with 
different size length. 
Concerning the zigzag patterns, we take advantage of the conservation of charge within one segment
that follows from the density-density nature of the interaction and 
the absence of the interorbital hopping in the undistorted system. One segment
of the z$n$ phase contains $n$ $a(b)$-orbital sites connected by 
the hopping $t$ and $n-2$ $b(a)$-orbital sites that contribute 
only through the occupation energy (and interlayer hopping if we deal with
the bilayer system). Thus, the problem of a z$n$-chain maps into that of interacting
molecules containing $2n-2$ pseudospins with $T=1/2$ where
the total $T^z$ is conserved in each molecule. For the presented calculations 
we take the following total numbers of segments $N$ for a given unit
length $n$, i.e., the pairs of $\{n,N\}$: $\{2,18\}$, $\{3,8\}$, $\{4,6\}$, 
$\{5,4\}$, $\{6,3\}$, $\{7,2\}$ and $\{8,2\}$. For the straight chain the 
number of $a$ and $b$-orbital sites is the same and equal to the chain
length $L$ that we assume to be $L=20$. The magnetizations in the 
zigzag segments and the occupations of the non-active orbitals are 
good quantum numbers in the problem. Hence, for each system we can directly determine 
the lowest energies in all the subspaces identified by the previous mentioned quantities.
All the systems are solved with periodic boundary conditions. We use the
translation invariance to reduce the total number of configurations. In case of the checkerboard
phases the approach is similar; checkerboard cell of the size $n\times n$
consists of $2n$ molecules of $n$ pseudospins $T=1/2$ interacting by a 
density-density interaction that conserves the total $T^z$. Both orbitals
are itinerant in the basic unit cell, thus the orbital occupations are not conserved quantities. 
The uniform FM phase is simulated by the largest checkerboard cell considered here,
which is $4\times 4$ assuming periodic boundary conditions.
Such methodology is particularly beneficial for obtaining the ground state 
of a given quantum problem without any specific bias in dealing with the interacting term. 
Furthermore, we expect that the 
finiteness of the system does not affect qualitatively the results because the noninteracting zigzag and
checkerboard states are insulating and the gap in the energy spectrum leads to short-range quantum correlations
that limit the size effects. Still, for the checkerboard configurations
the approach is particularly suited because we consider the charge as confined in each spin-polarized island
due to the large Hund coupling and only virtual processes can occur between neighbor blocks. 
On the other hand, the symmetry of the zigzag problem limits the Hilbert space for the 
computational analysis as even in the presence of the Coulomb interaction the
total charge is conserved within each zigzag unit. We focus the analysis on the monolayer 
system with tetragonal symmetry because it allows to simulate zigzag and checkerboard with larger size and 
it is more appropriate to capture the competition between short zigzag and the checkerboard 
states in the regime of doping close to x=0 where the effects of the interorbital interaction are
more relevant. This regime is indeed the case of dense interorbital occupation as
the number of minority spins for site is close to one.



\begin{references}
\bibitem{Ima98} M. Imada, A. Fujimori, and Y. Tokura,
                   Rev. Mod. Phys. \textbf{70}, 1039 (1998).
%
\bibitem{Dagotto2005} E. Dagotto, Science {\bf 309}, 257 (2005).
%
%
%
\bibitem{Kim2009} B.J. Kim, H. Ohsumi, T. Komesu, S. Sakai, T. Morita, 
				  H. Takagi, and T. Arima, Science {\bf 323}, 1329
				  (2009).
%
\bibitem{Pesin2010} D. Pesin and L. Balents, Nature Physics {\bf 6}, 376 (2010).
%
%

\bibitem{Moritomo} Y. Moritomo, Y. Tomioka, A. Asamitsu, Y. Tokura, and Y. Matsui, Phys. Rev. B {\bf 51}, 3297 (1995). 
%
\bibitem{Sternlieb} B.J. Sternlieb {\it et al.}, Phys. Rev. Lett. {\bf 76}, 2169 (1996).
%
\bibitem{Tokura} Y. Tokura, Rep. Prog. Phys. {\bf 69}, 797 (2006).
%
\bibitem{Tranquada} J.M. Tranquada, D.J. Buttrey, V. Sachan, and J.E. Lorenzo, Phys. Rev. Lett. {\bf 73}, 1003 (1994) 
%

%
\bibitem{Hotta2001} T. Hotta, A. Feiguin, and E. Dagotto, Phys. Rev. Lett. {\bf 86}, 4922 (2001).
%
\bibitem{Hotta2004} T. Hotta and E. Dagotto, Phys. Rev. Lett. {\bf 92}, 227201 (2004)
%
\bibitem{Dong2009} S. Dong, R. Yu, J.-M. Liu, and E. Dagotto, Phys. Rev. Lett. {\bf 103}, 107204 (2009).
%
\bibitem{Wrobel2010} P. Wr\'obel and A. M. Ole\'s, Phys. Rev. Lett. {\bf 104}, 206401 (2010).
%
\bibitem{Kruger2009} F. Kr${\rm{\ddot{u}}}$ger, S. Kumar, J. Zaanen, and J. van den Brink, Phys. Rev. B {\bf 79}, 054504 (2009).

\bibitem{Qi2012} T.F. Qi, O. B. Korneta, L. Li, K. Butrouna, V. S. Cao, Xiangang Wan, P. Schlottmann, R.K. Kaul, and G. Cao,
				 Phys. Rev. B {\bf 86}, 125105 (2012).
%
\bibitem{Cao2014} Y. Cao {\it et al.},
				  arXiv:1406.4978 (2014).
%
\bibitem{Lei2014} H. Lei, W.-G. Yin, Z. Zhong, and H. Hosono, Phys. Rev. B {\bf 89}, 020409(R) (2014).
%
\bibitem{Dhital2014} C. Dhital {\it et al.}, Nat. Comm.  {\bf 5}, 3377 (2014).
\bibitem{Brzezicki2015} W. Brzezicki, A.M. Ole\'s, and M. Cuoco, Phys. Rev. X {\bf 5}, 011037 (2015).


\bibitem{Ortmann2013} J.E. Ortmann, J.Y. Liu, J. Hu, M. Zhu, J. Peng, M. Matsuda, X. Ke, and Z.Q. Mao, Scientific Reports {\bf 3}, 2950 (2013). 


\bibitem{Mat2005} R. Mathieu, A. Asamitsu, Y. Kaneko, J. P. He, X. Z. Yu, R. Kumai, Y. Onose, N. Takeshita, T. Arima, H. Takagi, and Y. Tokura,
Phys. Rev. B {\bf 72}, 092404 (2005).


\bibitem{Hu2011} B. Hu, G. T. McCandless, V.O. Garlea, S. Stadler, Y. Xiong, J. Y. Chan, E. W. Plummer, and R. Jin,
Phys. Rev. B {\bf 84}, 174411 (2011).


\bibitem{Hos12} M.A. Hossain {\it et al.},
                   Phys. Rev. B \textbf{86}, 041102(R) (2012).

%
\bibitem{Pan2011} G. Panaccione {\it et al.},
New Journal of Physics {\bf 13}, 053059 (2011).

\bibitem{Mes12} D. Mesa, F. Ye, S. Chi, J.A. Fernandez-Baca,
                   W. Tian, B. Hu, R. Jin, E. W. Plummer, and J. Zhang,
                   Phys. Rev. B \textbf{85}, 180410(R) (2012).

\bibitem{Hos2013} M. A. Hossain {\it et al.},
Scientific Reports {\bf 3}, 2299 (2013).

\bibitem{Li2013} G. Li, Q. Li,
M. Pan, B. Hu, C. Chen, J. Teng, Z. Diao, J. Zhang, R. Jin, and 
E.W. Plummer,
Scientific Reports {\bf 3}, 2882 (2013).


\bibitem{DagottoReview} E. Dagotto, T. Hotta, and A. Moreo, Phys. Rep. {\bf 344}, 1 (2001).
%

%
\bibitem{Yunoki2000} S. Yunoki, T. Hotta, and E. Dagotto, Phys. Rev. Lett. {\bf 84}, 3714 (2000).
%

%
\bibitem{Hotta2003} T. Hotta, M. Moraghebi, A. Feiguin, A. Moreo, S. Yunoki, and E. Dagotto, Phys. Rev. Lett. \textbf{90}, 247203  (2003).
%
\bibitem{Kumar2010} S. Kumar, J. van den Brink, and A. P. Kampf, Phys. Rev. Lett. {\bf 104}, 017201 (2010).

\bibitem{Garcia2000} D.J. Garcia, K. Hallberg, C.D. Batista, M. Avignon, and B. Alascio, Phys. Rev. Lett. {\bf 85}, 3720 (2000).

\bibitem{Garcia2002} D.J. Garcia, K. Hallberg, C.D. Batista, S. Capponi, D. Poilblanc, M. Avignon, and B. Alascio, Phys. Rev. B {\bf 65}, 134444 (2002).


\bibitem{GarciaPRL2004} D.J. Garcia, K. Hallberg, B. Alascio, and M. Avignon, Phys. Rev. Lett. {\bf 93}, 177204 (2004).

\bibitem{Aliaga2001}  H. Aliaga, B. Normand, K. Hallberg, M. Avignon, and B. Alascio, Phys. Rev. B {\bf 64}, 024422 (2001).

\bibitem{Ye2009} F. Ye, S. Chi, J. A. Fernandez-Baca, A. Moreo, E. Dagotto, J. W. Lynn, R. Mathieu, Y. Kaneko, Y. Tokura, and P. Dai, Phys. Rev. Lett. {\bf 103}, 167202 (2009).
%


\bibitem{Medici2011} L. de Medici, J. Mravlje, and A. Georges, Phys. Rev. Lett. \textbf{107}, 256401 (2011).
%
\bibitem{Biermann2005} S. Biermann, L. de Medici, and A. Georges, Phys. Rev. Lett. \textbf{95}, 206401 (2005). 
%
%
\bibitem{Rincon2014} J. Rinc\'on, A. Moreo, G. Alvarez, and E. Dagotto, Phys. Rev. Lett. {\bf 112}, 106405 (2014).
%

%
\bibitem{supmat1} Details of the methodology, 
the numerical simulation and the expressions of the effective hopping in terms of the rotation and tilting angles are reported in the Supplemental Material.
%
%
\bibitem{Miyasaka2000} S. Miyasaka, T. Okuda, and Y. Tokura, Phys. Rev. Lett. {\bf 85}, 5388 (2000). 
%
\bibitem{Ishihara2005} S. Ishihara, Phys. Rev. Lett {\bf 94}, 156408 (2005).
%
\bibitem{Wohlfeld2006} K. Wohlfeld and A.M. Ole\'s,  Phys. Stat. Sol. (b) {\bf 243}, 142 (2006).
%
%
%
\bibitem{SlaterKoster} J.C. Slater and G.F. Koster, Phys. Rev. {\bf 94}, 1498 (1954).
%
%
\bibitem{Held2000} K. Held and D. Vollhardt, Phys. Rev. Lett. {\bf 84}, 5168 (2000).
%
\bibitem{Rozenberg1998} M. J. Rozenberg, Eur. Phys. J. B {\bf 2}, 457 (1998).
%
\bibitem{Golosov2010} D. I. Golosov, Phys. Rev. Lett. {\bf 104}, 207207 (2010).
%
%
\bibitem{Falicov1969} L. M. Falicov and J. C. Kimball, Phys. Rev. Lett. {\bf 22}, 997 (1969).



\end{references}
\end{document}